\documentclass[superscriptaddress,aps,pra,twocolumn,showpacs,nofootinbib,longbibliography]{revtex4-1}
\usepackage{amsmath}
\usepackage{latexsym}
\usepackage{amssymb}
\usepackage[colorlinks=true,citecolor=blue,urlcolor=blue]{hyperref}
\usepackage{color}
\usepackage{graphics,epstopdf}
\usepackage{soul}
\usepackage[demo]{graphicx}
\usepackage{subcaption}
\usepackage{lipsum}
\usepackage{algorithm2e}
\usepackage{physics}

\newcommand{\argp}[1]{\left( #1 \right)}

\begin{document}
\title{Device-Independent Quantum Secure Direct Communication Under Non-Markovian Quantum Channels}

\author{Pritam Roy}
\email{roy.pritamphy@gmail.com}
\affiliation{S. N. Bose National Centre for Basic Sciences, Block JD, Sector III, Salt Lake, Kolkata 700 098, India}

\author{Subhankar Bera}
\email{berasanu007@gmail.com}
\affiliation{S. N. Bose National Centre for Basic Sciences, Block JD, Sector III, Salt Lake, Kolkata 700 098, India}

\author{Shashank Gupta}
\email{shashankg687@gmail.com}
\affiliation{Okinawa Institute of Science and Technology Graduate University, Okinawa, Japan}
\affiliation{QuNu Labs Pvt. Ltd., M. G. Road, Bengaluru, Karnataka 560025, India}

\author{A. S. Majumdar}
\email{archan@bose.res.in}
\affiliation{S. N. Bose National Centre for Basic Sciences, Block JD, Sector III, Salt Lake, Kolkata 700 098, India}



\begin{abstract}
Device-independent quantum secure direct communication (DI-QSDC) is a promising primitive in quantum cryptography aimed towards addressing the problems of device imperfections and key management. However, significant effort is required to tackle practical challenges such as the distance limitation due to the decohering effects of quantum channels. Here, we explore the constructive effect of non-Markovian noise to improve the performance of DI-QSDC. Considering two different environmental dynamics modelled by the amplitude damping and the dephasing channels, we show that for both cases non-Markovianty leads to a considerable improvement over Markovian dynamics in terms of three benchmark performance criteria of the DI-QSDC task. Specifically, we find that non-Markovian noise (i) enhances the protocol security measured by Bell violation, (ii) leads to a lower quantum bit error rate, and (iii) enables larger communication distances by increasing the capacity of secret communication.
\end{abstract}

\maketitle

\section{Introduction} \label{1}
 In classical cryptography, secure communication is primarily based on the computational complexity of one-way mathematical functions \cite{Rivest78}. The computational complexity is reduced using quantum resources and algorithms like Shor's and Grover \cite{Shor94, Grover96, Long01} thus raising concern about the security of the traditional communication protocols in the post-quantum world. Quantum problems have quantum solutions giving emergence to quantum cryptography \cite{Gisin02}. Several quantum communication tasks have been proposed, including quantum teleportation \cite{bennet93}, quantum key distribution (QKD) \cite{Bennet92, BB84, ekert91}, quantum secure direct communication \cite{Long02}, and others \cite{Hillery}. The unique feature of detecting the presence of eavesdropper distinguishes quantum communication tasks from their classical counterparts. 

Quantum cryptography and quantum sensing \cite{Dengen17} are among the most advanced sectors of quantum technologies. Non-classical resources like entanglement \cite{ent}, steering \cite{steer} and Bell-nonlocality \cite{bell} are the primary driving forces for these technologies followed by advancements in the single photon sources and detectors \cite{Eisaman11}. Quantum key distribution is the most celebrated protocol under quantum cryptography with several successful trials and commercial products available currently \cite{Acin07, Pawlowski10, Pramanik, vazirani, Huang18, Tang19, Jan20, Farkas21, Jaskaran, Nadlinger21, Yash21, Bera23}. However, loophole-free self-testing, secure key management, migration and side channel attacks remain challenging hindrances \cite{Takahashi19}. 

Quantum secure direct communication (QSDC) has been put forward to address some of these challenges \cite{Long02}. In the realm of quantum communication, QSDC is currently a trending topic, both in theoretical \cite{Liu2022, Sheng2022, Ying2022, Cao2023, Zhao2024, Zhu2024, Zhang2024, Ahn2024, SGupta2023} and experimental applications \cite{WeiZhang, Wang2023, Qi2021, Zhang2022, Long2022, Paparelle2023, Pan20}. QSDC does not require key management, and hence, it is advantageous compared to quantum key distribution (QKD) and traditional classical communications. In recent years, various versions of QSDC have been proposed, such as single photon measurements based QSDC \cite{ZhangQ2023, Yang2020, Xiao2023}, measurement-device-independent QSDC \cite{Sun2023, Hong2023, Xiang2023, ZhouZR2020, Niu2020}, and continuous-variable QSDC \cite{Zhengwen2021}, and so on. However, component imperfections and implementation loopholes in realistic setups can compromise the security of the QSDC protocol as well. In this context, device-independent QSDC has been recently proposed \cite{LanZhou, LanZhou1, LanZhou23,Hong23} just like DI-QKD \cite{Acin07,Metger21,Xu21}. QSDC is regarded to have a huge potential for future communication networks and in developing a quantum internet \cite{Singh21,pan2023evolution}. 

One cannot escape the interaction with the ubiquitous environment during experimental or commercial realizations of such communication tasks. During this interaction,  quantum correlations, in general,  get destroyed, thereby creating obstacles in the successful implementation of the long-term execution of such tasks. The impact of the noise on quantum correlations has been studied extensively \cite{suddendeathent1,suddendeathent2,suddendeathent3,suddendeathent4,suddendeathent5,suddendeathent6,suddendeathent7}. Entanglement sudden death (ESD) may occur under interaction with dephasing channel noise. On the other hand, under certain circumstances,  a controlled environment may also aid in the preservation of quantum correlations \cite{Badziag2000, Bandyopadhyay02, GMN2006, Rivu21}. In particular, the revival of entanglement has been reported under non-Markovian channel interaction \cite{RLF17, nonMarkoent, Rivu22}. 

Techniques to mitigate the effect of noise on quantum correlations have been explored in the literature. For example, employing a scheme of weak measurements enables the preservation of various kinds of quantum correlations such as teleportation fidelity \cite{PM13}, quantum secret key rate \cite{DGPM2017},  and quantum non-bilocal correlations \cite{Gupta18} under the effect of the amplitude damping channel \cite{GGM21}. Noise can have a constructive effect on the teleportation fidelity and the coding capacity \cite{Rivu21}. More significantly, non-Markovian noise has been indicated to play a key role in the enhancement of quantum correlations in wide arenas such as quantum information \cite{nminfo, Rivu22} and metrology \cite{nmmetro}, prompting the development of resource theoretic frameworks for non-Markovianity \cite{Samya11, Samya12}. The development of advanced experimental technology may enable an open system to be driven from a Markovian to a non-Markovian regime \cite{nphys2011}. Furthermore, the role of memory effects on improving quantum communication protocols has been highlighted \cite{mele2023optical, PhysRevLett.129.180501, PhysRevA.106.042437}. 
 
 The above examples of constructive behaviour of the noise raises the question as to whether non-Markovian noise can improve the figure of merit for a general quantum communication task such as quantum secure direct communication ? We answer this question affirmatively in this work. In the present paper, we analyse the effect of the non-Markovian noise modelled by both the amplitude damping channel and the phase damping channel on the performance of a DI-QSDC task. In our protocol, the entangled pair of quantum particles interact with the channel noise. Through our analysis, we obtain a range of model parameters for which the non-Markovian noise enables better preservation of the nonclassical behaviour compared to the Markovian counterpart for the amplitude damping as well as dephasing channel noise.

The paper is organized as follows. In Sec. \ref{prelim}, we present a brief overview of the general DI-QSDC task along with the description of the open quantum systems and non-Markovian amplitude damping and dephasing quantum channels. In Sec. \ref{SS}, we consider a specific QSDC task where both the entangled quantum particles interact with the damping channel. In Sec. \ref{3}, we analyse the effect of the non-Markovian channels on the performance of the QSDC. Lastly, in Sec. \ref{Conclusion}, we present a summary of our
results and concluding remarks. 

\section{Preliminaries}{\label{prelim}}
In this Section, we will first recapitulate the notion of the QSDC. Then, we will consider the dynamics of open quantum systems and non-Markovian quantum channels under consideration for the present work.
\subsection{General QSDC protocol}
\begin{figure}[htbp]
\resizebox{8cm}{4.5cm}{\includegraphics{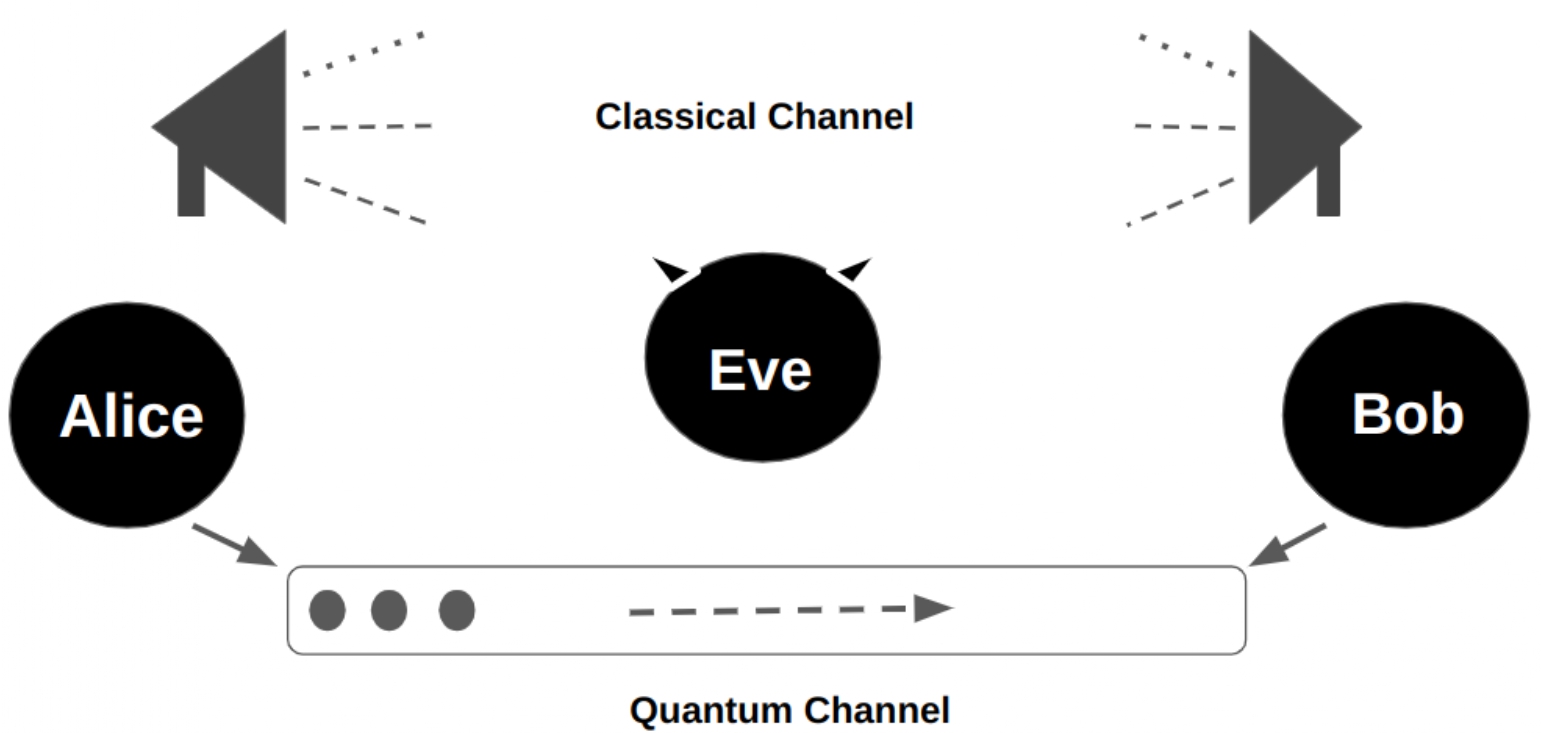}}
\caption{QSDC protocol.}
\label{model}
\end{figure}

QSDC facilitates the transmission of confidential messages through a quantum channel without the need for a pre-shared key. As shown in fig-(\ref{model}), authorized parties like Alice, the sender, encode messages using quantum states, and Bob, the receiver, decodes them through specific measurements and security checks. The security of this method relies heavily on the no-cloning theorem \cite{Wootters92} and the utilization of entanglement. Depending on the chosen non-classical resource, encoding and decoding scheme, QSDC can be either the prepare-measure type \cite{Deng04} or the entanglement-based type \cite{Deng03}. This approach offers a robust guarantee of message confidentiality, even when potential eavesdroppers like Eve are present, without the requirement of establishing a shared key in advance. QSDC harnesses the principles of quantum mechanics to ensure inherently secure communication, making it a highly promising technology for the future of secure data exchange.

\subsection{Dynamics of open quantum systems}
In contrast to isolated systems which undergo unitary evolution, the general quantum evolution of an open system can be described by a mathematical construct known as a Completely Positive Trace-Preserving (CPTP) map denoted as $\varepsilon(t_{\text{fin}}, t_{\text{ini}})$. This CPTP map functions as a transformation operator, mapping an initial density operator $\rho(t_{\text{ini}})$ from the set of density operators $(\mathcal{D}(\mathcal{H}))$ to a final density operator $\rho(t_{\text{fin}})$ within the same set. Importantly, we assume that an inverse operation, denoted as $\varepsilon^{-1}(t_{\text{fin}}, t_{\text{ini}})$, exists and is well-defined for all times ranging from $t_{\text{ini}}$ to $t_{\text{fin}}$. Consequently, we can express the dynamical evolution for any time interval $t_{\text{fin}}\geq t_{\text{int}} \geq t_{\text{ini}}$ as a composition,
\begin{equation}
    \varepsilon_{(t_{fin}, t_{ini})} = \varepsilon_{(t_{fin}, t_{int})}\circ \varepsilon_{(t_{int},t_{ini})}
    \label{C}
\end{equation}
The map $\varepsilon_{(t_{\text{fin}}, t_{\text{ini}})}$ is always completely positive as it represents a physical process. The map $\varepsilon_{(t_{\text{int}}, t_{\text{ini}})}$ is also completely positive due to the lack of interaction with the environment. However, $\varepsilon_{(t_{\text{fin}}, t_{\text{int}})}$ may not always be completely positive. A quantum dynamics applied to a system is considered divisible if it can be expressed in the form of equation \eqref{C} for any time interval $t_{\text{fin}}\geq t_{\text{int}} \geq t_{\text{ini}}$. Here, $t_{\text{ini}}$ denotes the initial time of the dynamics, and the symbol $\circ$ signifies the composition of two maps. 

This dynamics, denoted as $\varepsilon(t_{\text{fin}}, t_{\text{ini}})$, is termed positive divisible (P-divisible) when the map $\varepsilon(t_{\text{fin}}, t_{\text{int}})$ is positive for every $t_{\text{fin}}\geq t_{\text{int}} \geq t_{\text{ini}}$, satisfying the prescribed composition law. Likewise, the dynamics are termed completely positive divisible (CP-divisible) when the map $\varepsilon(t_{\text{fin}}, t_{\text{int}})$ is a Completely Positive Trace-Preserving (CPTP) map for every $t_{\text{fin}}\geq t_{\text{int}} \geq t_{\text{ini}}$ while adhering to the composition law.

The mathematical representation of a dynamical map $\varepsilon(t_{\text{fin}}, t_{\text{ini}})$ in terms of ``divisibility", where we describe memoryless evolution as a composition of physical maps, plays a pivotal role in defining the concept of quantum Markovianity. According to the RHP criterion  \cite{PhysRevLett.105.050403}, a dynamics is deemed non-Markovian if it fails to exhibit Complete Positive Divisibility (CP-divisibility). 
An alternative approach to characterizing non-Markovian dynamics, as proposed by Breuer et al. \cite{PhysRevLett.103.210401, PhysRevA.81.062115}, focuses on the distinguishability of quantum states after undergoing the action of a dynamical map. When a quantum system interacts with a noisy environment, the distinguishability between two quantum states gradually diminishes over time due to this interaction. However, if, at any given moment, the distinguishability of quantum states increases, it signifies a reverse flow of information from the environment back to the quantum system. This phenomenon serves as a clear indicator of non-Markovianity. 

The former method of identifying non-Markovian dynamics is referred to as ``RHP-type non-Markovianity" \cite{PhysRevLett.105.050403, rivas2014quantum}, while the latter is known as ``BLP-type non-Markovianity" \cite{PhysRevLett.103.210401, RevModPhys.88.021002}. 
It's important to note that a dynamics considered Markovian in the RHP sense will also be Markovian in the BLP sense. However, the reverse is not necessarily true in all cases. Hence, while the breaking of CP divisibility is a necessary condition, it is not always sufficient to conclude that there is information backflow from the environment to the quantum system.

\subsection{Non-Markovian Quantum channels} \label{4}
We will briefly explore some non-Markovian models, which are later employed to analyze the effectiveness of the DI-QSDC protocol. These models describe the behaviour of a system as it interacts with its environment, and this interaction is mathematically represented using Kraus operators,
\begin{equation}
    \varepsilon(\rho) = \sum_{i=1}^{n} E_{i}(t)\rho E_{i}^{\dagger}(t)
\end{equation}
In this context, we apply this method to characterize dissipative amplitude damping and pure dephasing interactions in the presence of non-Markovian environments. The Kraus operators that represent P-indivisible amplitude damping noise within the framework of non-Markovian effects \cite{bellomo2007non} on qubit systems can be expressed as follows:
\begin{equation}
    E_{0}^{AD} = \ket{0}\bra{0}+\sqrt{p_a(t)}\ket{1}\bra{1}\text{,}\;\; E_{1}^{AD} = \sqrt{{1-p_a(t)}}\ket{0}\bra{1}
    \label{AMP}
\end{equation}
where the function $p_a(t)$ is defined as $\exp{-\Gamma t} [\cos(b\frac{t}{2})+\frac{\Gamma}{b}\sin(b\frac{t}{2})]^{2}$, with $b = \sqrt{(2\gamma\Gamma -\Gamma^{2})}$. In these expressions, $\Gamma$ represents the linewidth, which is intricately linked to the reservoir's correlation time $\Gamma \approx \tau_{B}^{-1}$, while $\gamma$ signifies the coupling strength associated with the relaxation time of the qubit $\gamma \approx \tau_{R}^{-1}$. When the reservoir correlation time significantly exceeds the qubit relaxation time, we observe the emergence of memory effects. These effects are indicative of the non-Markovian nature of dissipation. Consequently, the relationship defined by $2\gamma/\Gamma$ dictates whether the noise displays Markovian characteristics when $2\gamma\leq\Gamma$ or non-Markovian behaviour otherwise.



Similarly,  a Kraus operator for the P-divisible dephasing channel can be expressed as \cite{utagi2020temporal}.
\begin{equation}
    E_{0}^{D_{ph}} = \ket{0}\bra{0}+p_d(t)\ket{1}\bra{1}\text{,}\;\; E_{1}^{D_{ph}} = \sqrt{{1-p^2_d(t)}}\ket{1}\bra{1}
    \label{DP}
\end{equation}
Here, the function $p_d(t)$ is defined as $\text{exp}[-\frac{\gamma}{2}\{t + \frac{1}{\Gamma}(\text{exp}(-\Gamma t)-1)\}]$. As the linewidth
$\Gamma$ tends towards infinity, the dephasing channel makes a transition towards the Markovian case.

\section{DI-QSDC protocol under environmental noise} \label{SS}

\begin{figure}[htbp]
\resizebox{8.5cm}{6.5cm}{\includegraphics{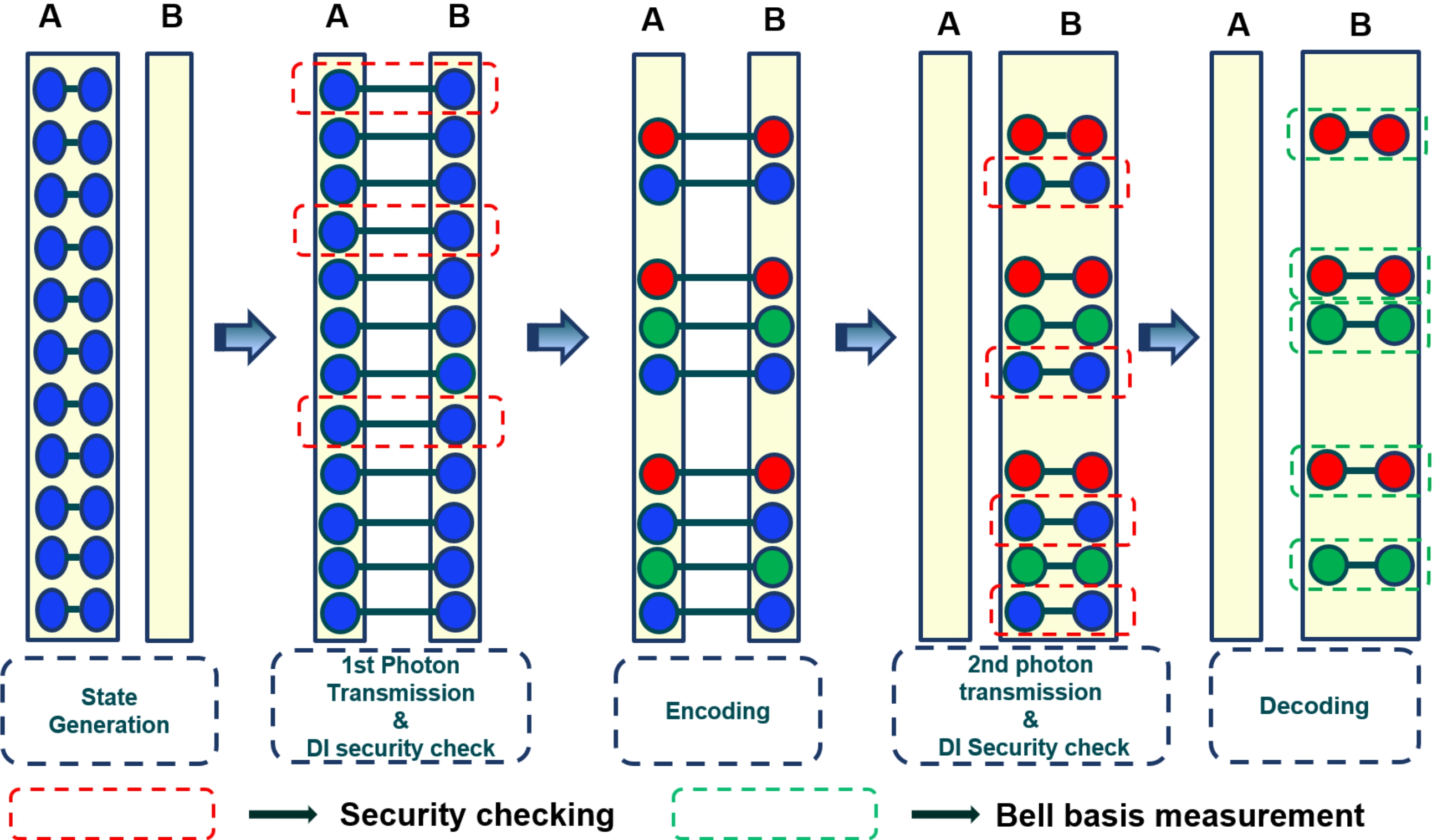}}
\caption{ Schematic diagram for QSDC protocol (Red colour entangled pair photons are $\ket{\psi^{+}}$, green colour entangled pair photons are $\ket{\psi^{-}}$ and blue colour entangled pair photons are $\ket{\phi^{+}}$).}
\label{QSDC1}
\end{figure}

The DI-QSDC protocol is based on two fundamental assumptions, first, unwanted information from Alice and Bob's physical surroundings may escape to the outer world \cite{Acin07}. The second is that quantum physics is true.
As shown in fig-(\ref{QSDC1}) the QSDC protocol is described in five stages \cite{LanZhou23}:

\textbf{Stage 1:} This stage is the state generation part, where Alice prepares N number of EPR pairs in her laboratory.  The checking (C) photon sequence and the message (M) photon sequence, are the two-photon sequences into which she separates the N EPR pairs. However, She will randomly tag the prepared EPR pairs as check or message pairs.

\textbf{Stage 2:} In this stage Alice uses the quantum non-Markovian channel to deliver Bob the photons in the C sequence one at a time. This process is called the first photon transmission. The single photon could lose its entire signal during transmission because of environmental noise and channel loss. The total photon transmission efficiency ($\eta$) directly depends upon the communication length between Alice and Bob ($D_{AB}$), which can be written as \cite{scaraniV},
\begin{equation}
\eta = 10^{-\frac{\alpha D_{AB}}{10}}
\label{PhotonT}
\end{equation}
where $\alpha$ is the attenuation factor. After the effect of decoherence, the initial state, $\rho_0=|\phi^+\rangle\langle\phi^+|$ becomes,
\begin{equation}
   \rho_{AB} =\sum_{i}(\mathbb{I}\otimes E_i)\rho_0(\mathbb{I}\otimes E^\dagger_i)
\end{equation}
where, $E_i$ are the Kraus operators defined in equation (\ref{AMP}) or equation (\ref{DP}), which satisfy $\sum_{i}E^\dagger_iE_i=\mathbb{I}$.
When both photon transmission loss and decoherence are considered, the N-shared states between Alice and Bob become,
\begin{equation}
    \rho_1=\eta\rho_{AB}+(1-\eta)\frac{\mathbb{I}}{4}
\label{1stPT}
\end{equation}

In this second stage, Alice and Bob also do the initial security checks to guarantee the safety of the first photon transmission operation. Alice first randomly chooses a large enough group of photons from the C photon sequence to perform the security check before revealing their positions to Bob over the public channel. Alice and Bob individually perform some measurements on each of the security-checking photons (C).
Alice has four possible binary measurements bases $A_0=\sigma_x$, $A_1=\frac{(\sigma_z+\sigma_x)}{\sqrt{2}}$, $A_2=\frac{(\sigma_z-\sigma_x)}{\sqrt{2}},$ and $A_3=\sigma_z$ with binary outcomes $a_i \in \lbrace -1,+1\rbrace$ and Bob has two possible measurement bases $B_0=A_0$ and $B_1=A_3$ with binary outcomes $b_j \in \lbrace -1,+1\rbrace$.
Once all of the checking photon pairs have been measured, Alice and Bob broadcast their measurement basis and results over a public channel.
There are four possibilities.

\textbf{Case 1:}If Alice choose $A_1$ or $A_2$ measurement bases then they can estimate the CHSH functional,
\begin{equation}
    S_l=\langle a_1b_1\rangle + \langle a_1b_2\rangle +\langle a_2b_1\rangle - \langle a_2b_2\rangle
    \label{S}
\end{equation}
where $l = 1, 2;$ for first and second round photon transmission respectively and $\langle a_ib_j\rangle = [P(a_i=b_j)-P(a_i\neq b_j)]$.
We assume that all measurement margins are random without losing generality, such that $\langle a_i\rangle=0, \langle b_j\rangle=0$ for $i\in \lbrace 0,1,2,3\rbrace$ and $j\in \lbrace 0,1\rbrace$.

\textbf{Case 2:} If Alice chooses measurement basis $A_0$ and Bob chooses measurement basis $B_0$ then the measurement result is used to estimate the quantum phase-flip error rate $Q_{p1}$ as,
\begin{equation}
    Q_{p1}=P(a_0\neq b_0)
\end{equation}

\textbf{Case 3:} If Alice chooses measurement basis $A_3$ and Bob chooses $B_1$, then the measurement result is used to estimate the quantum bit-flip error rate $Q_{b1}$ as,
\begin{equation}
    Q_{b1}=P(a_3\neq b_1)
\end{equation}
In this work, we are interested in the total Quantum Bit Error Rate (QBER),
\begin{equation}
  Q_l = Q_{pl}+Q_{bl}  
  \label{Q}
\end{equation}
where $l = 1, 2;$ for first and second round photon transmission respectively.
Note that entanglement-based QSDC uses four unitary operators in general to encode 2 bits of information, assuming that all four Bell states can be discriminated perfectly.
However, in the practical scenario of a linear optical set-up, the Bell state measurement (BSM) device can only distinguish two Bell states $\ket{\psi^{\pm}}$ \cite{LanZhou23}. In our analysis, we have used only two unitary operators to keep our findings aligned towards practical realization.

\textbf{Case 4:} The measurement results obtained when Alice choose $A_0$ and Bob choose $B_1$, or when Alice choose $A_3$ and Bob choose $B_0$, are ignored or removed from the remaining analysis. 

If the CHSH functional, $S_1\leq2$ then Alice and Bob are classically correlated. Since the first photon-transmission method is insecure in this scenario, the parties must terminate their communication because Eve might intercept the photons without being noticed. If the CHSH functional, $S_1>2$ then Alice and Bob are non-locally correlated. In this case, the rate at which Eve intercepts the photons is limited.
After making sure that the initial photon transfer is secure, Alice and Bob go on to the subsequent step.
  
\textbf{Stage 3:} This is the enoding part, where Alice retrieves the photons that were saved from the memory storage. She uses one of the two unitary procedures to encrypt her messages onto the message photon sequence (M). The two unitary operations are,
\begin{equation}
    U_0=\sigma_x, U_1=i\sigma_y
\end{equation}
Alice can change the state of the system from $|\phi^+\rangle$ to $|\psi^+\rangle$ and $|\psi^-\rangle$ using the unitary operations $U_0$ and $U_1$ respectively. Alice can encrypt her messages ``0" and ``1" onto the photon pairs by applying $U_0$ and $U_1$ respectively.

\textbf{Stage 4:} In the second photon transmission round, Alice chooses a few photons at random to serve as security check photons and she does not apply any operation on them. Alice deliberately shuffles the order of her photons in sequence and records the initial positions of each photon to prevent Eve from precisely intercepting the encoded photons based on her interceptions during the first round. Alice transmits all the photons from the sequence to Bob and publicly discloses the locations of each photon in the initial sequence after the transmission. Subsequently, Bob successfully reconstructs the original sequence. The second phase of security checking is subsequently carried out by Bob using Alice's announcement of the locations of the security checking photons.

The decoherence and the 2nd photon transmission loss influence the security-checking photon states on Bob's side. The security-checking photon state becomes, 
\begin{equation}
    \rho_2=\eta\Tilde{\rho}_{AB}+(1-\eta)\frac{\mathbb{I}}{4}
    \label{2ndPT}
\end{equation} 
 The state $\rho_
 {1}$ transforms to another mixed noisy state say,  $\Tilde{\rho}_{AB}
(=\sum_{i}(E_i \otimes \mathbb{I})\rho_{1}(E^\dagger_i \otimes \mathbb{I}))$,
due to transmission through the noisy channel.

After the measurements procedures, Bob can estimate the CHSH functional $S_2$ and calculate the $Q_{b2}$ and  $Q_{p_2}$. If the second photon-transmission procedure is not secure when $S_2\leq2$,  the parties abandon the communication. Otherwise, if the functional $S_2>2$ they ensure that the photon transmission is secure and go to the subsequent step.
It may be noted here that after the 2nd photon transmission, the CHSH function $S_1>S_2$ always.

\textbf{Stage 5:} In this final stage, Bob decodes the encrypted messages. For this, he performs the Bell basis measurements on all the remaining photons, and based on the measurements he can distinguish between $|\psi^+\rangle$ and $|\psi^-\rangle$, where the initial entangled state is $|\phi^+\rangle$.

At the end of the protocol, Alice and Bob calculate the secret message capacity and the maximum distance for which they can send the secret message.

In this device-independent scenario,  we make a general assumption that Eve obeys the laws of quantum physics. Furthermore, it is assumed that the measurement findings received by each side are entirely dependent on their present inputs.
Here we consider a collective attack in which Eve applies an identical attack against each of Alice's and Bob's systems. In this way,  all the photon
pairs have the same form after transmission. The ability to send secret messages $C_s$ is defined as the ratio of successfully and securely transferred qubits to the total number of encoded photon pairs. In scenarios involving non-ideal devices and noisy channels, the minimum achievable capacity for securely transmitting a secret message from Alice to Bob, considering collective attacks, is bounded below by the Devetak-Winter rate \cite{DW05},
\begin{equation}
    C_s\geq I(A:B)-I(A:E)
\end{equation}
where, $I(A:B)$  and $I(A:E)$ are the mutual information between Alice and Bob, and mutual information between Alice and Eve respectively. It is assumed that mutual information between Alice and Bob is uniformly marginally distributed, as \cite{Acin07,sp09},
 \begin{equation}
 \label{IAB}
     I(A:B)=1-h(Q_2)
 \end{equation}
where $Q_2$ $(= Q_{p2}+Q_{b2}$) is the total quantum bit error rate (QBER) after the 2nd photon transmission, and $h(p)$ is the binary entropy defined as,
\begin{equation}
\label{binary}
    h(p)=-p\log_2p-(1-p)\log_2(1-p)
\end{equation}
After the first and second rounds of photon transmission, we can estimate the Holevo quantities, given by,
\begin{align}
\label{chi}
    &\chi(S_1)\leq h\left(\frac{1+\sqrt{(S_1/2)^2-1}}{2}\right)
    \nonumber
    \\
    &\chi(S_2)\leq h\left(\frac{1+\sqrt{(S_2/2)^2-1}}{2}\right)
\end{align}
Comparing the Bell-CHSH functionals of the 1st round and 2nd round photon transmissions, one has $S_1>S_2$, and hence, it follows that the Holevo quantities must obey $\chi(S_1)<\chi(S_2)$.
\\

Here the mutual information between Alice and Eve $I(A:E)$ is denoted as the message intercepting rate, which is bounded by the Holevo quantity \cite{Hol73},
\begin{equation}
\label{IAE}
    I(A:E)\leq\chi(S_1)
\end{equation}
$I(A:E)$ can reach the maximum value of $\chi(S_1)$ only if, during the second round of photon transmission, Eve manages to intercept all the photons corresponding to those she intercepted in the initial photon transmission. However, since Alice reshuffles the sequence of her photons before the second transmission, the probability of $I(A:E)$ reaching $\chi(S_{1})$ diminishes significantly, particularly with a substantial number of transmitted photons.

Using the equations \eqref{IAB}, \eqref{chi} and \eqref{IAE} we can find the lower bound of $C_s$ to be,
\begin{equation}
\label{CCS}
    C_s\geq 1 - h(Q_{2})-\chi(S_1)
\end{equation}

By incorporating the Kraus operators from equation \eqref{AMP} into the expressions given by equations \eqref{1stPT} and \eqref{2ndPT}, we can determine the Bell-CHSH functional, denoted as $S_1$ and $S_2$ respectively, for both the photon transmissions. In the case of an amplitude damping channel, they are given by,
\begin{equation}
    S_{1} = \bigl( 1+ \sqrt{p_a(t)}\bigr)\eta\sqrt{2p_a(t)}\label{S1A}
\end{equation}
\begin{equation}
     S_{2} = \bigl( 1 - p_a(t) + 2p^2_a(t)\bigr)\eta^{2}\sqrt{2}\label{S2A}
\end{equation}
Likewise, by applying the Kraus operators from equation \eqref{DP}, $S_1$ and $S_2$, for  the  dephasing channel are given by,
\begin{equation}
    S_{1} = \bigl(1+p_d(t)\bigr)\eta\sqrt{2} 
    \label{S1D}
\end{equation}
\begin{equation}
    S_{2} = \bigl(1+p^2_d(t)\bigr)\eta^{2}\sqrt{2} 
\end{equation}
Next, using equations \eqref{1stPT} and \eqref{2ndPT} we can determine the total QBER in equation \eqref{Q}. $Q_1$ and $Q_2$ in the context of an amplitude damping channel are given by,
\begin{equation}
    Q_{1} = 1 - \frac{\eta\sqrt{p_a(t)}}{2}\bigl(1 + \sqrt{p_a(t)} \bigr)  
\end{equation}
\begin{equation}
    Q_{2} =  1 - \frac{\eta^{2}}{2}\bigl(1 - p_a(t) + 2p^2_a(t)\bigr)
    \label{Q2A}
\end{equation}
Similarly, the total QBER  in the case of the dephasing channel is given by,
\begin{equation}
    Q_{1} = 1 - \frac{\eta}{2}\bigl(1 + p_d(t) \bigr)
\end{equation}
\begin{equation}
    Q_{2} =  1 - \frac{\eta^{2}}{2}\bigr(1 + p^2_d(t) \bigl) 
    \label{Q2D}
\end{equation}
\begin{figure*}
    \begin{subfigure}{8.6cm}
\centering\includegraphics[width=8.6cm]{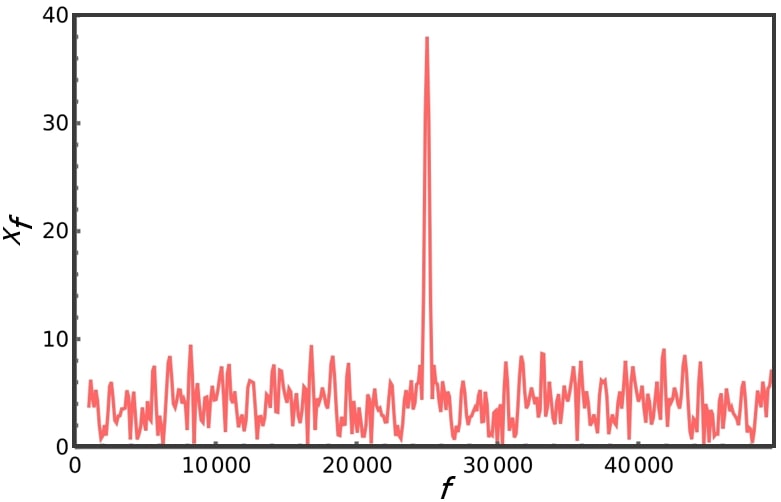}
    \caption{Spectrum determined by Bob. He concludes 25 KHz as modulation frequency and interprets Alice's message as $'0'$.}
    \label{125}
    \end{subfigure}
    \hspace{0.5 cm}
    \begin{subfigure}{8.6cm}
    \centering\includegraphics[width=8.6cm]{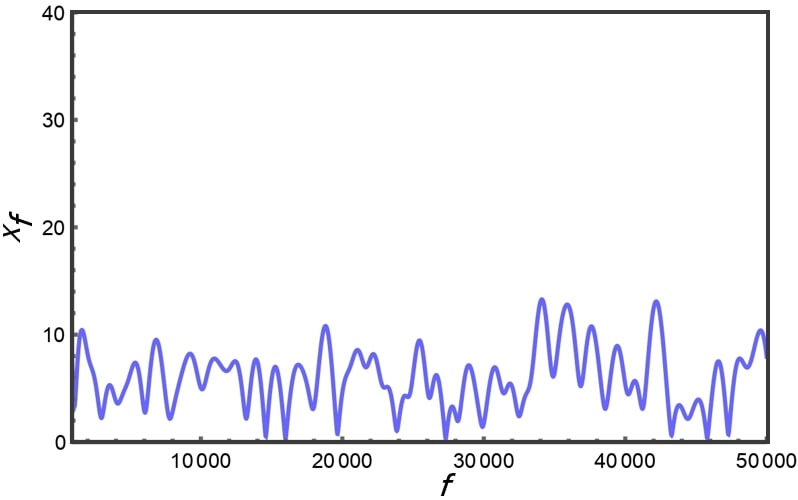}
    \caption{The spectrum of white noise is determined by an eavesdropper.}
    \label{25}
    \end{subfigure}
\caption{Discrete Fourier transform determined by Bob and an Eavesdropper when Alice chooses modulation signal at 25 KHz.}
\label{DFT_plot}
\end{figure*}

Now, from equations \eqref{binary}, \eqref{chi}, we get the secret message capacity $C_s$ under the amplitude damping channel, as,
 \begin{align}
    C_s &=  \Omega\log_{2}(\Omega)+(1-\Omega)\log_{2}(1-\Omega)\nonumber\\
    &+\frac{1-\delta}{2}\log_{2}(1-\delta)+\frac{1+\delta}{2}\log_{2}(1+\delta)
 \end{align}
 where, $\Omega=\frac{\eta^2}{2}(1-p_a(t)+2p^2_a(t))$, 
 and $\delta=\sqrt{\frac{\eta^2p_a(t)}{2}(1+\sqrt{p_a(t)})^2-1}$.
 
For the dephasing channel,  we compute the secret message capacity to be,
\begin{align}
    C_s &=  \omega\log_{2}(\omega)+(1-\omega)\log_{2}(1-\omega)\nonumber\\
    &+\frac{1-\Delta}{2}\log_{2}(1-\Delta)+\frac{1+\Delta}{2}\log_{2}(1+\Delta)
 \end{align}
 where, $\omega=\frac{\eta^2}{2}(1+p^2_d(t))$, 
 and $\Delta=\sqrt{\frac{\eta^2}{2}(1+p_d(t))^2-1}$.\\
 
 In contrast to quantum key distribution,  QSDC transmits secret messages directly rather than random keys. Hence, the parties are unable to use the post-processing approach to fix message errors or loss. In the case of encoding at a single photon level, one can quantify the effect of loss and noise on the rate of information transmission. The loss rate of information ($r_{lr}$) is defined as the number of lost information qubits divided by the total amount of information qubits. The error rate of information ($r_{er}$) is calculated as the number of wrong qubits read out by Bob divided by the total number of information qubits that Bob may read. Under the both noisy channels  $r_{lr}$ and $r_{er}$ are calculated as,
 \begin{equation}
     r_{lr}= 1-\eta^2
     \nonumber
 \end{equation}
 \begin{equation}
     r_{er}= \frac{1}{2}\argp{1-p^2_{a(d)}(t)}
 \end{equation}
\begin{figure*}
    \begin{subfigure}{8.6 cm}
    \centering\includegraphics[width=8.6 cm]{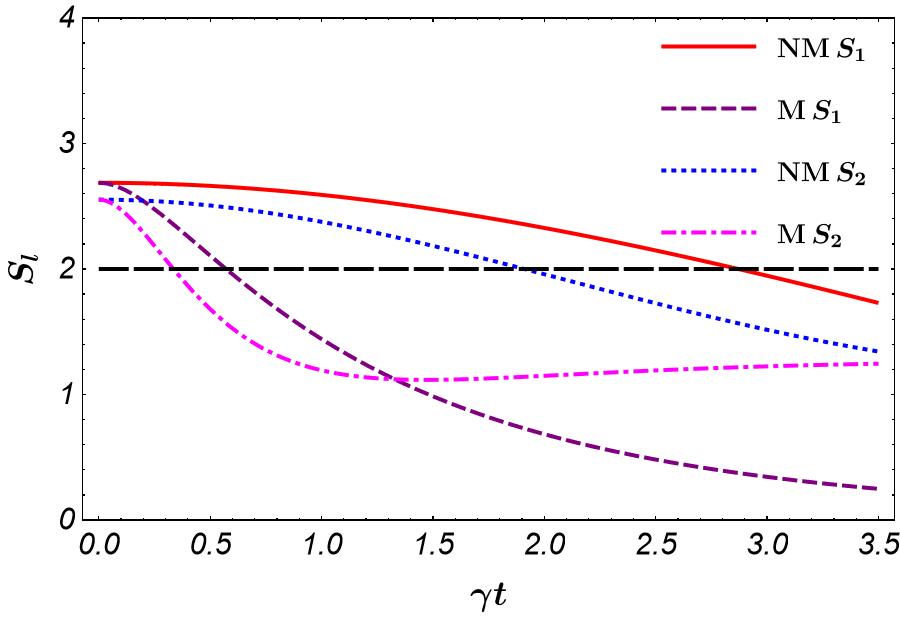}
    \caption{ The Bell-CHSH functional ($\boldsymbol{S_l}$) (where $\boldsymbol{l} = 1, 2 $ for first and second photon transmission respectively) is plotted against Noise parameter ($\boldsymbol{\gamma t}$) for Markovian (\textbf{M}) ($\Gamma = 5\gamma$) and non-Markovian (\textbf{NM}) ($\Gamma = 0.1\gamma$) amplitude damping channel (here $\eta = 0.95 $).}
    \label{SA}
    \end{subfigure}
    \hspace{0.5 cm}
    \begin{subfigure}{8.6 cm}
    \centering\includegraphics[width=8.6 cm]{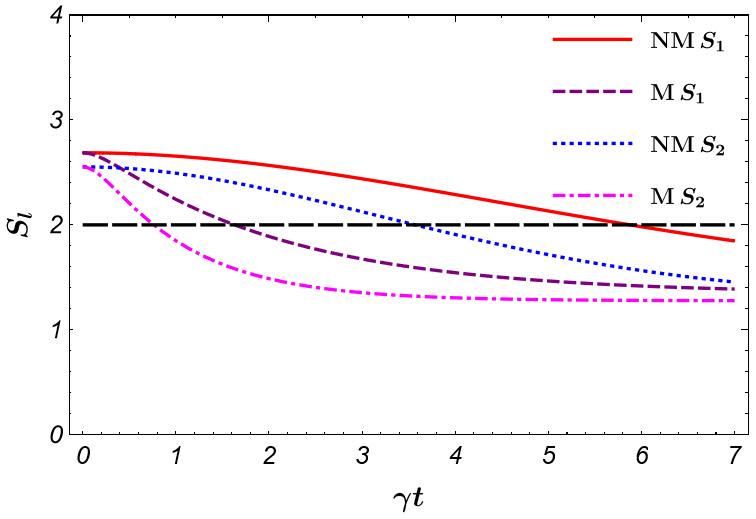}
    \caption{ The Bell-CHSH functional ($\boldsymbol{S_l}$) (where $\boldsymbol{l} = 1, 2$ for first and second photon transmission respectively) is plotted against Noise parameter ($\boldsymbol{\gamma t}$) for Markovian (\textbf{M}) ($\Gamma = 5\gamma$) and non-Markovian (\textbf{NM}) ($\Gamma = 0.1\gamma$) dephasing channel (here $\eta = 0.95 $).}
    \label{SD}
    \end{subfigure}
\caption{The Bell-CHSH functional (for both round photon transmissions) is plotted against the noise parameter ($\boldsymbol{\gamma t}$) for the amplitude damping and dephasing channels for the Markovian and non-Markovian regions.}
\label{graphs1}
\end{figure*}
\begin{figure*}

    \begin{subfigure}{8.6 cm}
    \centering\includegraphics[width=8.6 cm]{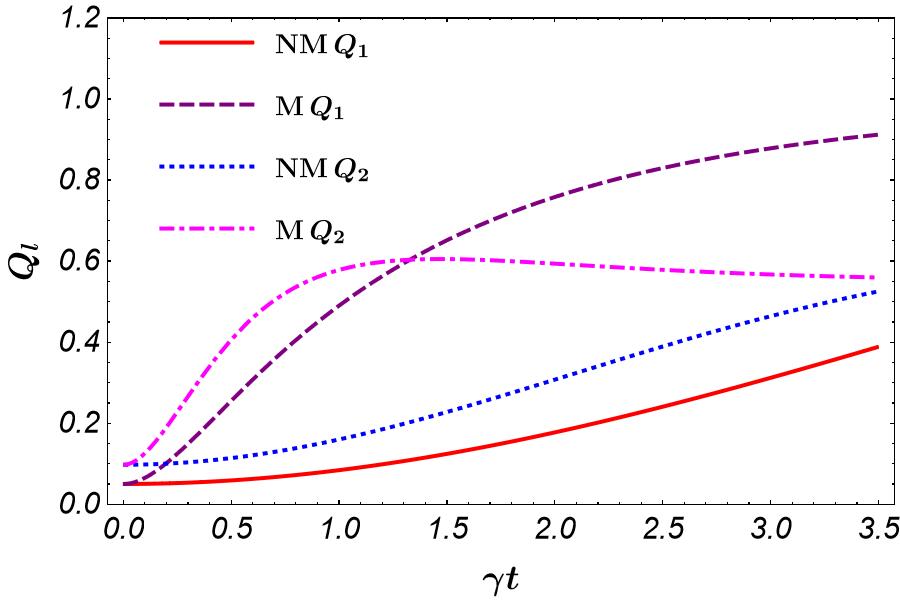}
    \caption{ The QBER ($\boldsymbol{Q_l}$) (where $\boldsymbol{l} = 1, 2$ for first and second photon transmission respectively) is plotted against Noise parameter ($\boldsymbol{\gamma t}$) for Markovian (\textbf{M}) ($\Gamma = 5\gamma$) and non-Markovian (\textbf{NM}) ($\Gamma = 0.1\gamma$) amplitude damping channel (here $\eta = 0.95 $).}
    \label{QA}
    \end{subfigure}
    \hspace{0.5cm}
    \begin{subfigure}{8.6 cm}
    \centering\includegraphics[width=8.6 cm]{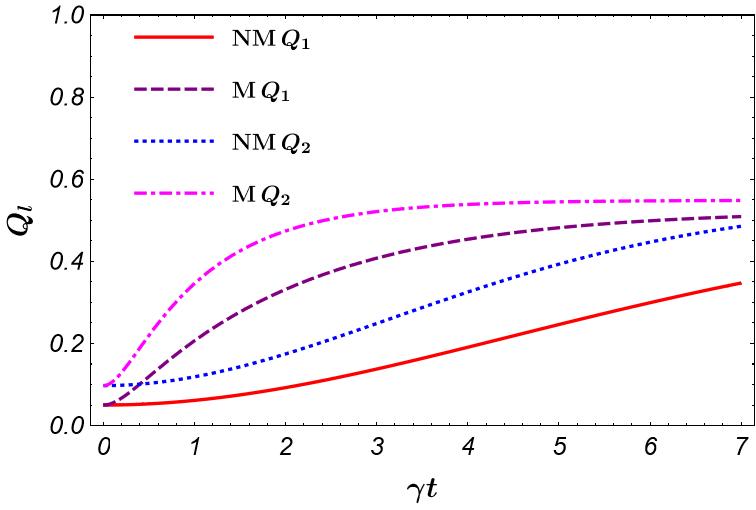}
    \caption{ The QBER ($\boldsymbol{Q_l}$) (where $\boldsymbol{l} = 1, 2$ for first and second photon transmission respectively) is plotted against Noise parameter ($\boldsymbol{\gamma t}$) for Markovian (\textbf{M}) ($\Gamma = 5\gamma$) and non-Markovian (\textbf{NM}) ($\Gamma = 0.1\gamma$) dephasing channel (here $\eta = 0.95 $).}
    \label{QD}
    \end{subfigure}%
\caption { The quantum bit error rate (QBER) (for both round photon transmissions) is plotted against the noise parameter ($\boldsymbol{\gamma t}$) for the amplitude damping and dephasing channels for the Markovian and non-Markovian regions.}
\label{graphs2}
\end{figure*}
\begin{figure*}
    \begin{subfigure}{8.6cm}
    \centering\includegraphics[width=8.6cm]{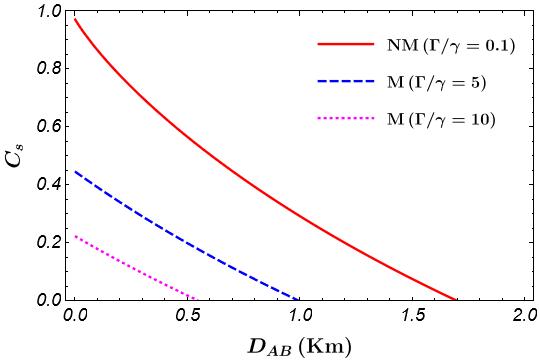}
    \caption{ The secret message capacity ($\boldsymbol{C_{s}}$) is plotted against Communication length ($\boldsymbol{D_{AB}}$ (\textbf{Km})) for Markovian (\textbf{M}) ($\Gamma/\gamma = 5, 10$) and non-Markovian (\textbf{NM}) ($\Gamma/\gamma = 0.1$) amplitude damping channel for $\gamma t = 0.15$.}
    \label{CSA}
    \end{subfigure}%
    \hspace{0.5cm}
    \begin{subfigure}{8.6cm}
    \centering\includegraphics[width=8.6cm]{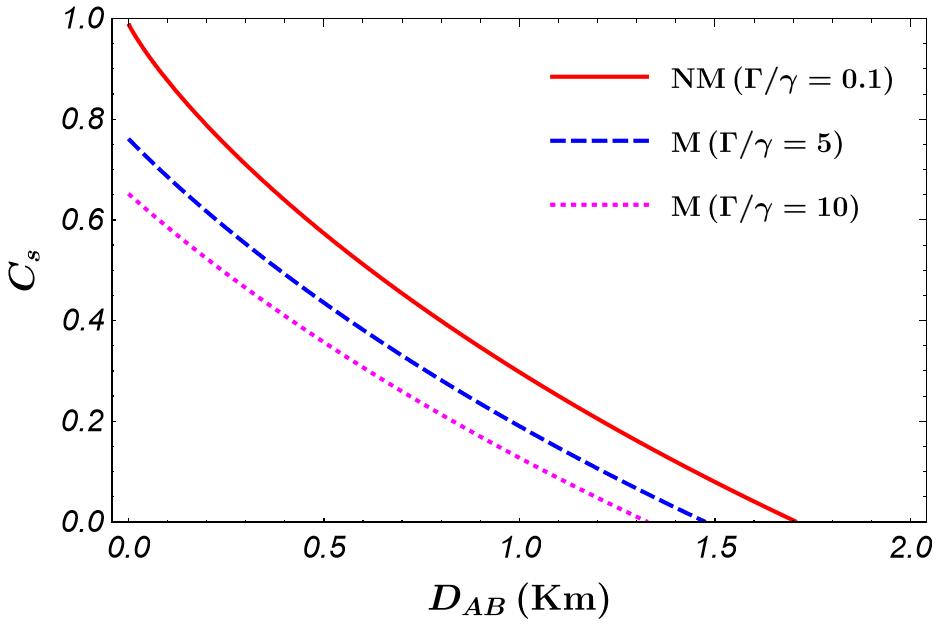}
    \caption{ The secret message capacity ($\boldsymbol{C_{s}}$) is plotted against Communication length ($\boldsymbol{D_{AB}}$ (\textbf{Km})) for Markovian (\textbf{M}) ($\Gamma/\gamma = 5, 10$) and non-Markovian (\textbf{NM}) ($\Gamma/\gamma = 0.1$) dephasing channel for $\gamma t = 0.15$.}
    \label{CSD}
    \end{subfigure}
\caption{The secret message capacity ($\boldsymbol{C_{s}}$) is plotted against communication length ($\boldsymbol{D_{AB}}$ (\textbf{Km})) for amplitude damping and dephasing channels for the Markovian and non-Markovian regions.}
\label{graphs3}
\end{figure*}

 This indicates that the rate of information decreases with both the increasing loss and error implying the inefficiency of the encoding at the single-photon level. In order to handle this aspect, the block encoding technique may be employed as follows \cite{Hu2016}:

Instead of using an individual photon to encode a bit value, block encoding applies a series of
operations periodically (say 1 milli second) on a single-photon block (say 10000 photons) to encode information. Alice applies unitary operation in the block according to a periodic function with period $T=\frac{1}{f}$, where $f$ is the modulation frequency that encodes the information.
Bob can reliably decode the modulation frequency using Discrete-time Fourier transform logic by determining the Bell state measurement result. He performs the decoding operation only if the QBER at stage 4 is within the threshold. The modulation frequency used for the block can be determined as follows:
\begin{equation}
    X_{(f)} = \sum_{i=1}^N x_{(i)}e^{-j.2\pi f \tau_i}.
\label{DFT}
\end{equation}
where $x_{(i)} = 0$ or $1$ depending upon whether the Bell state measurement is $\ket{\psi^+}$ or $\ket{\psi^-}$. $N$ is the number of detections (say $100$ after 20 dB channel loss). $'f'$ is an element from the range of modulation frequencies. $\tau_i$ is the detection time-stamp within the $100$ photon block \cite{Hu2016}. Let us understand this with an example:

Suppose Alice wants to send a message $'0110'$. She uses a modulation frequency of 25 KHz for $'0'$ bit and 50 KHz for $'1'$ bit. By modulation frequency, we mean the rate at which Alice will be applying the unitary operator ($U_0$ and $U_1$) on the 10000 photon block. So, for this specific message, Alice's modulation frequency choice is 25 KHz, 50 KHz, 50 KHz, and 25 KHz respectively. We now elaborate the block encoding and decoding procedure at 25 KHz:

$\bullet$ \textbf{Block encoding:} In a 1 milli second block duration, Alice's modulation signal contains 50 pulses and she applies unitary operation $U_0$ when the modulation signal is high and $U_1$ when the modulation signal is high on the 10000 photons, respectively.

$\bullet$ \textbf{Bob's decoding:} Bob performs Bell state measurement on the received photons (say 100). He now has $x_{(i)}$ and $\tau_i \quad \forall i \in \{1, 100\}$. He calculates $X_{(f)}$ using Eq. (\ref{DFT}) for frequency range ($0, 50$ 
 KHz) and interprets Alice's modulation frequency where $X_{(f)}$ is maximum. Any outside party (Eavesdropper) cannot determine $x_{(i)}$ and so will end up with a white noise spectrum as shown in fig-(\ref{DFT_plot}).

 In the following section, we compute the three key performance indicators of the DI-QSDC protocol, namely, the communication capacity, the quantum bit error rate and the Bell-CHSH functional under the action of Markovian and non-Markovian amplitude damping as well as dephasing noise channels 
choosing values of parameters in an experimentally realizable range \cite{tang2012, passos2019}.

\section{Effect of non-markovian channels} \label{3}
In this section, we discuss the dynamical behaviour of the benchmark physical parameters when the DI-QSDC task is performed between Alice and Bob under the action of a noisy channel and attack by an eavesdropper. To evaluate the effect of non-Markovian dynamics, the ratio $\Gamma/\gamma$ plays the most significant role. For instance, transitions between the Markovian and non-Markovian regimes have been realized in cavity quantum electrodynamics experimental configurations \cite{kuhr2007ultrahigh} using Rydberg atoms with a lifetime of $\tau_{\text{at}} = 30$ ms,  with a cavity lifetime of $\tau_{\text{cav}} = 130$ ms. The single photon's relaxation time in a superconducting cavity is $25.6 \pm 0.2$. ms\cite{milul2023superconducting}. In our subsequent calculations, the ratio of the $\gamma$ and $\Gamma$ parameters are chosen to lie within the experimentally feasible range of all-optical set-ups \cite{tang2012, passos2019}. Additionally, we take the attenuation factor to be $\alpha = 0.2$ dB/Km, which corresponds to a transmission
efficiency $\eta \approx 95\%$ for a communication distance of the order
of $1$ Km \cite{zapatero2019}.

Let us first consider stage 2, when Alice and Bob perform the DI security checking task after the first photon transmission.  In fig-(\ref{SA}) we show the dynamics of the Bell-CHSH functionals ($S_1$, $S_2$) against the noise parameter $\gamma t$. Here we consider the amplitude damping channel. It can be seen that we get a larger non-classical region (up to $\gamma t$ = 3 (red)) under non-Markovian dynamics compared to the non-classical region  (up to $\gamma t$ = 0.6 (purple)) under Markovian dynamics. We next consider stage 4, when the second DI security checking task is performed after the second photon transmission. In this case, too, compared to a Markovian regime (up to $\gamma t$ = 0.4 (magenta)), a non-Markovian regime (up to $\gamma t$ = 2 (blue)) corresponds to a bigger non-classical region.

A similar behaviour is exhibited for the dephasing channel. As seen from fig-(\ref{SD}) when the dephasing channel is taken into account, here too we get a larger non-classical region in the non-Markovian regime (up to $\gamma t$ = 5.8 (red) and 3.6 (blue)) compared to the Markovian regime (up to $\gamma t$ = 1.6 (purple) and 0.8 (magenta) after the 1st and 2nd photon transmission, respectively) in the behaviour of the Bell-CHSH functionals ($S_1$, $S_2$). 
\\
It is worthwhile to note that for the same value of the noise parameter $\gamma$, a non-Markovian channel leads to a higher value of the Bell-CHSH parameter, 
signifying an enhanced level of security compared to a Markovian channel. Comparing fig-(\ref{SA}) and fig-(\ref{SD}), it may be noted further
that the dephasing channel enables the achievement of a bigger non-classical region compared to the case of the amplitude damping channel, a result
that is valid irrespective of the Markovian or non-Markovian nature of
the dynamics.

We next analyze the behaviour of the quantum bit error rate under various
dynamics considered here. In figs-(\ref{QA}, \ref{QD}) where we plot total QBER ($Q_1$, $Q_2$) (after the first and second photon transmissions, respectively) against the noise parameter $\gamma t$. It can be seen that we get less QBER in the non-Markovian regime compared to the  Markovian regime under both amplitude and dephasing noise. Thus, the action of a non-Markovian channel leads to a lesser error rate than that under the action of a Markovian channel for
both the amplitude damping and the dephasing noise. Further, a comparison of fig-(\ref{QA}) and fig-(\ref{QD}) exhibits less QBER value under dephasing noise. One can see that QBER values can be more than $0.8$ under amplitude damping noise, but it does not exceed the value of $0.5$ when we consider dephasing noise. So, it turns out that as with the case of the Bell-CHSH functional, a dephasing channel turns out to be a better option for our DI-QSDC protocol compared to an amplitude damping channel for either Markovian or non-Markovian dynamics.

We finally investigate the secret message capacity $C_s$.
From fig-(\ref{CSA}) in the case of amplitude damping channel, we can observe that the secret message can be communicated over a maximum communication distance of $D_{AB} = 1.7$ km (red) in the non-Markovian regime. This gives an advantage over the Markovian channel, where we can communicate only up to $D_{AB} = 0.55$ km (magenta) for $\Gamma/\gamma = 10$.  After increasing the strength of the parameter but still confining it to the Markovian
regime, one sees that $D_{AB}$ increases to 1 km (blue). Further, one can see 
that the the secret message capacity can be maximum, {\it i.e.} almost
$1$, in the non-Markovian regime for very short distances, while in the Markovian regime, it is considerably lesser.

In fig-(\ref{CSD}) we display the secret message capacity under the action of dephasing noise. Here too the advantage under non-Markovian dynamics is
exhibited, corresponding to a larger communication distance compared to
the case of Markovian dynamics. Similarly again,  secret message capacity 
at short distances is considerably larger under non-Markovian dynamics. 
If we compare the maximum communication under non-Markovian channels for
the amplitude damping with the dephasing channel (comparing fig-(\ref{CSA}) and fig-(\ref{CSD}), it is observed that both turn out to be nearly the same.
In contrast, the maximum communication length is larger for the dephasing
noise undergoing Markovian dynamics compared to the amplitude-damping
dynamics.

\section{Conclusions} \label{Conclusion}

Quantum cryptography is a rapidly developing sector under quantum technologies. The advancement in single photon sources and detectors has accelerated interest in quantum key distribution products. However, three primary challenges need to be addressed: (i) decoherence due to interaction with the channel noise, (ii) device imperfections and implementation loopholes, and (iii) security challenges such as authentication, key management and key migration. In this regard, device-independent quantum communication tasks have been proposed that somewhat address the issues (ii) and (iii). Considerable further effort is required to address the issue (i). 

With the above motivation in the present work, we analyse the performance of device-independent quantum secure direct communication under non-Markovian noise modelled by amplitude damping and dephasing quantum channels.
Here we have performed a detailed analysis of the effect of non-Markovian noise on the efficacy of the DI-QSDC task when both the entangled photons interact with the decohering quantum channel one at a time during their transit in the DI-QSDC protocol. We have investigated the role of non-Markovian noise in Bell-inequality violation, quantum bit error rate (QBER), and communication capacity.  We have also incorporated the idea of block encoding to address the decreasing rate of information at the single-photon level encoding due to loss and error.

Our results show that non-Markovian environmental dynamics lead to enhanced Bell violation, a decrease in the qubit error rate, as well an increase in the communication capacity. Improvement of performance of the DI-QSDC protocol is thus displayed concerning all these three benchmark parameters for both in the case of non-Markovian amplitude damping as well as non-Markovian dephasing quantum channels. Our present analysis should motivate further theoretical studies on different quantum communication protocols under realistic environmental scenarios modelled by Markovian and non-Markovian noisy channels, and also experimental reservoir engineering protocols aimed towards driving
open systems from the Markovian to the non-Markovian regime \cite{nphys2011, wang2018}.

{\it Acknowledgements:} SB and ASM acknowledge support from the Project
No. DST/ICPS/QuEST/2018/98 from the Department
of Science and Technology, Government of India. SG
acknowledges the support from QuNu Labs Pvt Ltd and OIST, Japan.


\begin{thebibliography}{99}
\bibitem{Rivest78} 
Rivest, R. L., Shamir, A., Adleman, L.: A method for obtaining digital signatures and public-key cryptosystems. Communications of the ACM \textbf{21}(2), 120--126 (1978).
\url{https://doi.org/10.1145/359340.359342}

\bibitem{Shor94} 
Shor, P. W.: Algorithms for quantum computation: discrete logarithms and factoring. 
in Proceedings 35th Annual Symposium on Foundations of Computer Science,
Santa Fe, NM, USA, 1994, pp. 124-134.
\url{https://doi.org/10.1109/SFCS.1994.365700}

\bibitem{Grover96} 
Grover, L. K.: A fast quantum mechanical algorithm for database search. arXiv:quant-ph/9605043 (1996). \url{https://doi.org/10.48550/arXiv.quant-ph/9605043}

\bibitem{Long01} 
Long, G. L.: Grover algorithm with zero theoretical failure rate. Phys. Rev. A \textbf{64}, 022307 (2001).
\url{https://doi.org/10.1103/PhysRevA.64.022307}

\bibitem{Gisin02} 
Gisin, N., Ribordy, G., Tittel, W., Zbinden, H.: Quantum cryptography. Rev. Mod. Phys. \textbf{74}, 145 (2002).
\url{https://doi.org/10.1103/RevModPhys.74.145}

\bibitem{bennet93} 
Bennett, C. H., Brassard, G., Cr\'epeau, C., Jozsa, R., Peres, A., Wootters, W. K.: Teleporting an unknown quantum state via dual classic and Einstein-Podolsky-Rosen channels. Phys. Rev. Lett. \textbf{70}, 1895–1899 (1993).
\url{https://doi.org/10.1103/PhysRevLett.70.1895}

\bibitem{Bennet92} 
Bennett, C. H.: Quantum cryptography using any two nonorthogonal states. Phys. Rev. Lett. \textbf{68}, 3121 (1992).
\url{https://doi.org/10.1103/PhysRevLett.68.3121}

\bibitem{BB84} 
Bennett, C. H., Brassard, G.: Quantum cryptography, public key distribution and coin tossing. Theoretical Computer Science \textbf{560} (Part 1), 7-11 (2014).
\url{https://doi.org/10.1016/j.tcs.2014.05.025}

\bibitem{ekert91} 
Ekert, A. K.: Quantum cryptography based on Bell's theorem. Phys. Rev. Lett. \textbf{67}, 661 (1991).
\url{https://doi.org/10.1103/PhysRevLett.67.661}


\bibitem{Long02} 
Long, G. L., Liu, X. S.: Theoretically efficient high-capacity quantum-key-distribution scheme. Phys. Rev. A \textbf{65}, 032302 (2002).
\url{https://doi.org/10.1103/PhysRevA.65.032302}

\bibitem{Hillery} 
Hillery, M., Bu\v{z}ek, V., Berthiaume, A.: Quantum secret sharing. Phys. Rev. A \textbf{59}, 1829 (1999).
\url{https://doi.org/10.1103/PhysRevA.59.1829}

\bibitem{Dengen17} 
Degen, C. L., Reinhard, F., Cappellaro, P.: 
Quantum sensing. Rev. Mod. Phys. \textbf{89}, 035002 (2017).
\url{https://doi.org/10.1103/RevModPhys.89.035002}

\bibitem{ent} 
Horodecki, R., Horodecki, P., Horodecki, M., Horodecki, K.: Quantum entanglement. Rev. Mod. Phys \textbf{81}, 865 (2009).
\url{https://doi.org/10.1103/RevModPhys.81.865}

\bibitem{steer} 
Uola, R., Costa, Ana C. S., Nguyen, H. Chau, Gühne, O.: Quantum steering. Rev. Mod. Phys. \textbf{92}, 015001 (2020).
\url{https://doi.org/10.1103/RevModPhys.92.015001}

\bibitem{bell} 
Brunner, N., Cavalcanti, D., Pironio, S., Scarani, V., Wehner, S.: Bell nonlocality. Rev. Mod. Phys. \textbf{86}, 419 (2014).
\url{https://doi.org/10.1103/RevModPhys.86.419}

\bibitem{Eisaman11} 
Eisaman, M. D., Fan, J., Migdall, A., Polyakov, S. V.: Invited Review Article: Single-photon sources and detectors. Rev. Sci. Instrum. \textbf{82}, 071101 (2011).
\url{https://doi.org/10.1063/1.3610677}

\bibitem{Acin07} 
Acín, A., Brunner, N., Gisin, N., Massar, S., Pironio, S., Scarani, V.: Device-Independent Security of Quantum Cryptography against Collective Attacks. Phys. Rev. Lett. \textbf{98}, 230501 (2007).
\url{https://doi.org/10.1103/PhysRevLett.98.230501}

\bibitem{Pawlowski10} 
Pawłowski, M.: 
Security proof for cryptographic protocols based only on the monogamy of Bell’s inequality violations. Phys. Rev. A \textbf{82}, 032313 (2010).
\url{https://doi.org/10.1103/PhysRevA.82.032313}

\bibitem{Pramanik} 
Pramanik, T., Kaplan, M., Majumdar, A. S.: Fine-grained Einstein-Podolsky-Rosen steering inequalities. Phys. Rev. A \textbf{90}, 050305(R) (2014).
\url{https://doi.org/10.1103/PhysRevA.90.050305}

\bibitem{vazirani} 
Vazirani, U., Vidick, T.: Fully Device Independent Quantum Key Distribution. Phys. Rev. Lett. \textbf{113}, 140501 (2014).
\url{https://doi.org/10.1103/PhysRevLett.113.140501}

\bibitem{Huang18} 
Khan, I., Heim, B., Neuzner, A., Marquardt, C.: 
Satellite-based qkd. Opt. Photonics News \textbf{29}(2), 26-33 (2018).
\url{https://doi.org/10.1364/OPN.29.2.000026}

\bibitem{Tang19} 
Tang, B. Y., Liu, B., Zhai, Y. P., et al.: High-speed and Large-scale Privacy Amplification Scheme for Quantum Key Distribution. Sci Rep \textbf{9}, 15733 (2019).
\url{https://doi.org/10.1038/s41598-019-50290-1}

\bibitem{Jan20} 
Kołodyński, J., Máttar, A., Skrzypczyk, P., Woodhead, E., Cavalcanti, D., Banaszek, K., Acín, A.: Device-independent quantum key distribution with single-photon sources. Quantum \textbf{4}, 260 (2020).
\url{https://doi.org/10.22331/q-2020-04-30-260}

\bibitem{Farkas21} 
Farkas, M., Juandó, M. B., Łukanowski, K., Kołodyński, J., Acín, A.: Bell Nonlocality Is Not Sufficient for the Security of Standard Device-Independent Quantum Key Distribution Protocols. Phys. Rev. Lett. \textbf{127}, 050503 (2021).
\url{https://doi.org/10.1103/PhysRevLett.127.050503}

\bibitem{Jaskaran} 
Singh, J., Ghosh, S., Arvind, Goyal, S. K.: Role of Bell-CHSH violation and local filtering in quantum key distribution. Physics Letters A, Volume \textbf{392}, 127158 (2021).
\url{https://doi.org/10.1016/j.physleta.2021.127158}

\bibitem{Nadlinger21} 
Nadlinger, D. P., Drmota, P., Nichol, B. C., Araneda, G., Main, D., Srinivas, R., Lucas, D. M., Ballance, C. J., Ivanov, K., Tan, E. Y.-Z., Sekatski, P., Urbanke, R. L., Renner, R., Sangouard, N., Bancal, J.-D.: Experimental quantum key distribution certified by Bell's theorem. Nature \textbf{607}, 682–686 (2022).
\url{https://doi.org/10.1038/s41586-022-04941-5}

\bibitem{Yash21} 
Wath, Y., Hariprasad, M., Shah, F., Gupta, S.:  Eavesdropping a Quantum Key Distribution network using sequential quantum unsharp measurement attacks. Eur. Phys. J. Plus \textbf{138}, 54 (2023).
\url{https://doi.org/10.1140/epjp/s13360-023-03664-4}

\bibitem{Bera23} 
Bera, S., Gupta, S., Majumdar, A. S.: Device-independent quantum key distribution using random quantum states. Quantum Inf Process \textbf{22}, 109 (2023).
\url{https://doi.org/10.1007/s11128-023-03852-2}

\bibitem{Takahashi19} 
Takahashi, R., Tanizawa, Y., Dixon, A.: 
A high-speed key management method for quantum key distribution network, 
in Eleventh International Conference on Ubiquitous and Future Networks (ICUFN), Zagreb, Croatia, 2019, pp. 437-442.
\url{https://ieeexplore.ieee.org/document/8806052}


\bibitem{Liu2022} 
Liu, X., Luo, D., Lin, G. L., Chen, Z. H., Huang, C. F., Li, S. Z., Zhang, C. X., Zhang, Z. R., Wei, K. J.:
Fiber-based quantum secure direct communication without active polarization compensation.
Sci. China Phys. Mech. Astron. \textbf{65} (12), 120311 (2022).
\url{https://doi.org/10.1007/s11433-022-1976-0}

\bibitem{Sheng2022} 
Sheng, Y. B., Zhou, L., Long, G. L.:
One-step quantum secure direct communication.
Sci. Bull. (Beijing) \textbf{67} (4), 367 (2022).
\url{https://doi.org/10.1016/j.scib.2021.11.002}

\bibitem{Ying2022} 
Ying, J. W., Zhou, L., Zhong, W., Sheng, Y. B.:
Measurement-device-independent one-step quantum secure direct communication.
Chin. Phys. B \textbf{31} (12), 120303 (2022).
\url{https://doi.org/10.1088/1674-1056/ac8f37}


\bibitem{Cao2023} 
Cao, Z. W., Lu, Y., Chai, G., Yu, H., Liang, K. X., Wang, L.:
Realization of Quantum Secure Direct Communication with Continuous Variable.
Research \textbf{6}, 0193 (2023).
\url{https://doi.org/10.34133/research.0193}

\bibitem{Zhao2024} 
Zhao, P., Zhong, W., Du, M. M., Li, X. Y., Zhou, L., Sheng, Y. B.:
Quantum secure direct communication with hybrid entanglement.
Frontiers of Physics \textbf{19}, 51201 (2024).
\url{https://doi.org/10.1007/s11467-024-1396-5}

\bibitem{Zhu2024} 
Zhu, P. H., Zhong, W., Du, M. M., Li, X. Y., Zhou, L., Sheng, Y. B.:
One-step quantum dialogue.
Chin. Phys. B \textbf{33}, 030302 (2024).
\url{https://doi.org/10.1088/1674-1056/ad1c5c}

\bibitem{Zhang2024} 
Zhang, C., Zhou, L., Zhong, W., Du, M. M., Sheng, Y. B.: Measurement-device-independent quantum dialogue based on entanglement swapping and phase encoding.
Quantum Inf Process \textbf{23}, 52 (2024). \url{https://doi.org/10.1007/s11128-024-04260-w}

\bibitem{Ahn2024} 
Ahn, B., Park, J., Lee, J., Lee, S.:
High-dimensional single photon based quantum secure direct communication using time and phase mode degrees.
Sci. Rep. \textbf{14}, 888 (2024).
\url{https://doi.org/10.1038/s41598-024-51212-6}

\bibitem{SGupta2023}
Gupta, S.: Experimental simulation of the quantum secure direct communication using MATLAB and Simulink.
Eur. Phys. J. Plus \textbf{138}, 913 (2023).
\url{https://doi.org/10.1140/epjp/s13360-023-04532-x}


\bibitem{WeiZhang} 
Zhang, W., Ding, D. S., Sheng, Y. B., Zhou, L., Shi, B. S., Guo, G. C.: Quantum Secure Direct Communication with Quantum Memory. Phys. Rev. Lett. \textbf{118}, 220501 (2017).
\url{https://doi.org/10.1103/PhysRevLett.118.220501}

\bibitem{Wang2023} 
Wang, M., Zhang, W., Guo, J. X., Song, X. T., Long, G. L.:
Experimental demonstration of secure relay in quantum secure direct communication network.
Entropy \textbf{25} (11), 1548 (2023).
\url{https://doi.org/10.3390/e25111548}


\bibitem{Qi2021} 
Qi, Z., Li, Y., Huang, Y., Feng, J., Zheng, Y., Chen, X.: A 15-user quantum secure direct communication network.
Light Science \& Applications \textbf{10}, 183 (2021).
\url{https://doi.org/10.1038/s41377-021-00634-2}

\bibitem{Zhang2022} 
Zhang, H., Sun, Z., Qi, R., Yin, L., Long, G. L., Lu, J.:
Realization of quantum secure direct communication over 100 km fiber with time-bin and phase quantum states.
Light Science \& Applications \textbf{11}, 83 (2022).
\url{https://doi.org/10.1038/s41377-022-00769-w}

\bibitem{Long2022} 
Long, G.-L., Pan, D., Sheng, Y.-B., Xue, Q., Lu, J., Hanzo, L.:
An Evolutionary Pathway for the Quantum Internet Relying on Secure Classical Repeaters.
IEEE Network \textbf{36} (3), 82-88 (2022).
\url{https://doi.org/10.1109/MNET.108.2100375}

\bibitem{Paparelle2023} 
Paparelle, I., Mousavi, F., Scazza, F., Bassi, A., Paris, M., Zavatta, A.:
Practical quantum secure direct communication with squeezed states. arXiv:2306.14322 [quant-ph] (2023). \url{https://doi.org/10.48550/arXiv.2306.14322}

\bibitem{Pan20} 
Pan, D., Lin, Z., Wu, J., Zhang, H., Sun, Z., Ruan, D., Yin, L., Long, G. L.: Experimental free-space quantum secure direct communication and its security analysis. Photonics Res. \textbf{8}, 1522–1531 (2020).
\url{https://doi.org/10.1364/PRJ.388790}

\bibitem{ZhangQ2023} 
Zhang, Q., Du, M. M., Zhong, W., Sheng, Y. B., Zhou, L.: Single-Photon Based Three-Party Quantum Secure Direct Communication with Identity Authentication. Ann. Phys. (Berlin), \textbf{536}, 2300407 (2023).
\url{https://doi.org/10.1002/andp.202300407}

\bibitem{Yang2020} 
Yang, L., Wu, J., Lin, Z., Yin, L., Long, G.:
Quantum secure direct communication with entanglement source and single-photon measurement.
Sci. China Phys. Mech. Astron. \textbf{63}, 110311 (2020).
\url{https://doi.org/10.1007/s11433-020-1576-y}

\bibitem{Xiao2023} 
Xiao, Y.-X., Zhou, L., Zhong, W., Du, M.-M., Sheng, Y.-B.: The hyperentanglement-based quantum secure direct communication protocol with single-photon measurement. Quantum Inf Process \textbf{22}, 339 (2023).
\url{https://doi.org/10.1007/s11128-023-04097-9}


\bibitem{Sun2023} 
Sun, Z. Z., Pan, D., Ruan, D., Long, G. L.: 
One-Sided Measurement-Device-Independent Practical Quantum Secure Direct Communication. Journal of Lightwave Technology \textbf{41}, no. 14, 4680-4690 (2023).
\url{https://doi.org/10.1109/JLT.2023.3244880}

\bibitem{Hong2023} 
Hong, Y.-P., Zhou, L., Zhong, W., Sheng, Y.-B.: Measurement-device-independent three-party quantum secure direct communication. Quantum Inf Process \textbf{22}, 111 (2023).
\url{https://doi.org/10.1007/s11128-023-03853-1}

\bibitem{Xiang2023} 
Li, X.-J., Pan, D., Long, G.-L., Hanzo, L.: Single-Photon-Memory Measurement-Device-Independent Quantum Secure Direct Communication—Part II: A Practical Protocol and its Secrecy Capacity. IEEE Communications Letters, vol. \textbf{27}, no. 4, pp. 1060-1064 (2023).
\url{https://doi.org/10.1109/LCOMM.2023.3247176}

\bibitem{ZhouZR2020}
Zhou, Z. R., Sheng, Y. B., Niu, P. H., Yin, L. G., Long, G. L., Hanzo, L.:
Measurement-device-independent quantum secure direct communication.
\textit{Sci. China Phys. Mech. Astron.} \textbf{63}, 230362 (2020).
\url{https://doi.org/10.1007/s11433-019-1450-8}
\bibitem{Niu2020} 
Niu, P. H., Wu, J. W., Yin, L. G., Long, G. L.:
Security analysis of measurement-device-independent quantum secure direct communication.
Quantum Inf Process \textbf{19}, 356 (2020).
\url{https://doi.org/10.1007/s11128-020-02840-0}

\bibitem{Zhengwen2021} 
Cao, Z., Wang, L., Liang, K., Chai, G., Peng, J.: Continuous-Variable Quantum Secure Direct Communication Based on Gaussian Mapping. Phys. Rev. Applied \textbf{16}, 024012 (2021).
\url{https://doi.org/10.1103/PhysRevApplied.16.024012}

\bibitem{LanZhou} 
Zhou, L., Sheng, Y. B., Long, G. L.:
Device-independent quantum secure direct communication against collective attacks.
Science Bulletin, Volume \textbf{65}, 12-20 (2020).
\url{https://doi.org/10.1016/j.scib.2019.10.025}


\bibitem{LanZhou1} 
Zhou, L., Sheng, Y. B.: One-step device-independent quantum secure direct communication. Sci. China Phys. Mech. Astron. \textbf{65}, 250311 (2022).
\url{https://doi.org/10.1007/s11433-021-1863-9}


\bibitem{LanZhou23} 
Zhou, L., Xu, B.-W., Zhong, W., Sheng, Y.-B.: Device-Independent Quantum Secure Direct Communication with Single-Photon Sources. Phys. Rev. Applied \textbf{19}, 014036 (2023).
\url{https://doi.org/10.1103/PhysRevApplied.19.014036}


\bibitem{Hong23} 
Zeng, H., Du, M.-M., Zhong, W., Zhou, L., Sheng, Y.-B.: High-capacity device-independent quantum secure direct communication based on hyper-encoding. Fundamental Research (2023).
\url{https://doi.org/10.1016/j.fmre.2023.11.006}

\bibitem{Metger21} 
Metger, T., Dulek, Y., Coladangelo, A., Friedman, R. A.: Device-independent quantum key distribution from computational assumptions. New J. Phys. \textbf{23}, 123021 (2021).
\url{https://doi.org/10.1088/1367-2630/ac304b}


\bibitem{Xu21} 
Xu, F., Zhang, Y. Z., Zhang, Q., Pan, J.: 
Device-independent quantum key distribution with random post selection. Phys. Rev. Lett. \textbf{128}, 110506 (2022).
\url{https://doi.org/10.1103/PhysRevLett.128.110506}

\bibitem{Singh21} 
Singh, A., Dev, K., Siljak, H., Joshi, H. D., Magarini, M.: Quantum Internet—Applications, Functionalities, Enabling Technologies, Challenges, and Research Directions.
IEEE Communications Surveys and Tutorials. vol. \textbf{23}, no. 4, pp. 2218-2247, Fourth quarter 2021.
\url{https://doi.org/10.1109/COMST.2021.3109944}

\bibitem{pan2023evolution} 
Pan, D., Long, G.-L., Yin, L., Sheng, Y.-B., Ruan, D., Ng, S. X., Lu, J., Hanzo, L.: 
The Evolution of Quantum Secure Direct Communication: On the Road to the Qinternet. 
IEEE Communications Surveys \& Tutorials (Early Access).
\url{https://doi.org/10.1109/COMST.2024.3367535}

\bibitem{suddendeathent1} 
    Yu, T., Eberly, J. H.: Finite-Time Disentanglement Via Spontaneous Emission. Phys. Rev. Lett. \textbf{93}, 140404 (2004). \url{https://doi.org/10.1103/PhysRevLett.93.140404}
\bibitem{suddendeathent2}
    Dodd, P.J., Halliwell, J. J.: Disentanglement and decoherence by open system dynamics. Phys. Rev. A \textbf{69}, 052105 (2004). \url{https://doi.org/10.1103/PhysRevA.69.052105}
\bibitem{suddendeathent3}    
    Yu, T., Eberly, J. H.: Quantum Open System Theory. Bipartite Aspects Phys. Rev. Lett. \textbf{97}, 140403 (2006). \url{https://doi.org/10.1103/PhysRevLett.97.140403}
\bibitem{suddendeathent4}    
    Almeida, M. P., de Melo, F., Hor-Meyll, M., Salles, A., Walborn, S. P., Ribeiro, P.H. S., Davidovich, L.: Environment-Induced Sudden Death of Entanglement. Science \textbf{316}, 579 (2007). \url{https://doi.org/10.1126/science.1139892}
\bibitem{suddendeathent5}    
    Bellomo, B., Franco, R. L., Maniscalco, S., Compagno, G.: Entanglement trapping in structured environments. Phys. Rev. A \textbf{78}, 060302(R) (2008). \url{https://doi.org/10.1103/PhysRevA.78.060302}
\bibitem{suddendeathent6}    
    Salles, A., de Melo, F., Almeida, M.P., Hor-Meyll, M., Walborn, S. P., Ribeiro, P.H. S., Davidovich, L.: Experimental investigation of the dynamics of entanglement: Sudden death, complementarity, and continuous monitoring of the environment. Phys. Rev. A \textbf{78}, 022322 (2008). \url{https://doi.org/10.1103/PhysRevA.78.022322}
\bibitem{suddendeathent7}    
    Yu, T., Eberly, J. H.: Sudden Death of Entanglement. Science \textbf{323}, 598 (2009). \url{https://doi.org/10.1126/science.1167343}

\bibitem{Badziag2000} 
Badziag, P., Horodecki, M., Horodecki, P., Horodecki, R.: Local environment can enhance fidelity of quantum teleportation. Phys. Rev. A \textbf{62}, 012311 (2000).
\url{https://doi.org/10.1103/PhysRevA.62.012311}

\bibitem{Bandyopadhyay02} 
Bandyopadhyay, S.: Origin of noisy states whose teleportation fidelity can be enhanced through dissipation. 
Phys. Rev. A \textbf{65}, 022302 (2002).
\url{https://doi.org/10.1103/PhysRevA.65.022302}

\bibitem{GMN2006} 
Ghosh, B., Majumdar, A. S., Nayak, N.: 
Environment assisted entanglement enhancement. Phys. Rev. A \textbf{74}, 052315 (2006).
\url{https://doi.org/10.1103/PhysRevA.74.052315}

\bibitem{Rivu21} 
Gupta, R., Gupta, S., Mal, S., Sen De, A.: 
Performance of Dense Coding and Teleportation for Random States: Augmentation via Pre-processing, Phys. Rev. A \textbf{103}, 032608 (2021).
\url{https://doi.org/10.1103/PhysRevA.103.032608}

\bibitem{RLF17} Franco, R. L., Compagno, G.. Lectures on General Quantum Correlations and their Applications, in Lectures on General Quantum Correlations and their Applications, edited by F. F. Fanchini, D. O. S. Pinto, and G. Adesso, Springer, 2017. \url{https://doi.org/10.1007/978-3-319-53412-1}

\bibitem{nonMarkoent} Mazzola, L., Maniscalco, S., Piilo, J., Suominen, K.-A., Garraway, B. M.: Sudden death and sudden birth of entanglement in common structured reservoirs. Phys. Rev. A, \textbf{79}, 042302 (2009). \url{https://doi.org/10.1103/PhysRevA.79.042302}

\bibitem{Rivu22} 
Gupta, R., Gupta, S., Mal, S., Sen De, A.:
Constructive Feedback of Non-Markovianity on Resources in Random Quantum States.
Phys. Rev. A \textbf{105}, 012424 (2022).
\url{https://doi.org/10.1103/PhysRevA.105.012424}

\bibitem{PM13} 
Pramanik, T., Majumdar, A. S.:
Improving the fidelity of teleportation through noisy channels using weak measurement.
Phys. Lett. A \textbf{377}, 3209 (2013).
\url{https://doi.org/10.1016/j.physleta.2013.10.012}

\bibitem{DGPM2017} 
Datta, S., Goswami, S., Pramanik, T., Majumdar, A. S.: Preservation of a lower bound of quantum secret key rate in the presence of decoherence.
Phys. Lett. A \textbf{381}, 897 (2017).
\url{https://doi.org/10.1016/j.physleta.2017.01.019}

\bibitem{Gupta18} 
Gupta, S., Datta, S., Majumdar, A. S.:
Preservation of quantum non-bilocal correlations in noisy entanglement-swapping experiments using weak measurements.
Phys. Rev. A \textbf{98}, 042322 (2018).
\url{https://doi.org/10.1103/PhysRevA.98.042322}

\bibitem{GGM21} 
Goswami, S., Ghosh, S., Majumdar, A. S.:
Protecting quantum correlations in the presence of amplitude damping channel.
Phys. A: Math. Theor. \textbf{54}, 045302 (2021).
\url{https://doi.org/10.1088/1751-8121/abd59f}

\bibitem{nminfo} 
Laine, E.-M., Breuer, H.-P., Piilo, J.: Nonlocal memory effects allow perfect teleportation with mixed states.
Scientific Reports \textbf{4}, 4620 (2014).
\url{https://doi.org/10.1038/srep04620}

\bibitem{nmmetro} 
Altherr, A., Yang, Y.: Quantum Metrology for Non-Markovian Processes.
Phys. Rev. Lett. \textbf{127}, 060501 (2021).
\url{https://doi.org/10.1103/PhysRevLett.127.060501}

\bibitem{Samya11} 
Bhattacharya, S., Bhattacharya, B., Majumdar, A. S.:
Thermodynamic utility of non-Markovianity from the perspective of resource interconversion.
J. Phys. A: Math. Theor. \textbf{53}, 335301 (2020).
\url{https://doi.org/10.1088/1751-8121/aba0ee}
\bibitem{Samya12} 
Bhattacharya, S., Bhattacharya, B., Majumdar, A. S.:
Convex resource theory of non-Markovianity.
J. Phys. A: Math. Theor. \textbf{54}, 035302 (2020).
\url{https://doi.org/10.1088/1751-8121/abd191}

\bibitem{nphys2011} 
Liu, B.-H., Li, L., Huang, Y., Li, C.-F., Guo, G.-C., Laine, E.-M., Breuer, H.-P., Piilo, J.: Experimental control of the transition from Markovian to non-Markovian dynamics of open quantum systems.
Nature Physics \textbf{7}, 931 (2011).
\url{https://doi.org/10.1038/nphys2085}


\bibitem{mele2023optical} 
Mele, F. A., De Palma, G., Fanizza, M., Giovannetti, V., Lami, L.: Optical fibres with memory effects and their quantum communication capacities, arXiv:2309.17066 [quant-ph] (2023). \url{https://doi.org/10.48550/arXiv.2309.17066}

 \bibitem{PhysRevLett.129.180501} 
Mele, F. A., Lami, L., Giovannetti, V.: Restoring Quantum Communication Efficiency over High Loss Optical Fibers.
Phys. Rev. Lett. \textbf{129}, 180501 (2022).
\url{https://doi.org/10.1103/PhysRevLett.129.180501}

 \bibitem{PhysRevA.106.042437} 
Mele, F. A., Lami, L., Giovannetti, V.: Quantum optical communication in the presence of strong attenuation noise.
Phys. Rev. A \textbf{106}, 042437 (2022).
\url{https://doi.org/10.1103/PhysRevA.106.042437}

\bibitem{Wootters92} 
Wootters, W., Zurek, W.:
A single quantum cannot be cloned. Nature, \textbf{299}, 802–803 (1982). \url{https://doi.org/10.1038/299802a0}.

\bibitem{Deng04} 
Deng, F. G., Long, G. L.: Secure direct communication with a quantum one-time pad.
Phys. Rev. A \textbf{69}, 052319 (2004).
\url{https://doi.org/10.1103/PhysRevA.69.052319}

\bibitem{Deng03} 
Deng, F. G., Long, G. L., Liu, X. S.:
Two-step quantum direct communication protocol using the Einstein-Podolsky-Rosen pair block.
Phys. Rev. A \textbf{68}, 042317 (2003).
\url{https://doi.org/10.1103/PhysRevA.68.042317}

\bibitem{PhysRevLett.105.050403} 
Rivas, A., Huelga, S. F., Plenio, M. B.: Entanglement and Non-Markovianity of Quantum Evolutions.
Phys. Rev. Lett. \textbf{105}, 050403 (2010).
\url{https://doi.org/10.1103/PhysRevLett.105.050403}

\bibitem{PhysRevLett.103.210401} 
Breuer, H.-P., Laine, E.-M., Piilo, J.: Measure for the Degree of Non-Markovian Behavior of Quantum Processes in Open Systems.
Phys. Rev. Lett. \textbf{103}, 210401 (2009).
\url{https://doi.org/10.1103/PhysRevLett.103.210401}

\bibitem{PhysRevA.81.062115} 
Laine, E.-M., Piilo, J., Breuer, H.-P.: Measure for the non-Markovianity of quantum processes. Phys. Rev. A \textbf{81}, 062115 (2010).
\url{https://doi.org/10.1103/PhysRevA.81.062115}


\bibitem{rivas2014quantum} 
Laine, E.-M., Piilo, J., Breuer, H.-P.: Quantum non-Markovianity: characterization, quantification and detection, Reports on Progress in Physics \textbf{77}, 094001 (2014). \url{https://doi.org/10.1088/0034-4885/77/9/094001}
 
\bibitem{RevModPhys.88.021002} 
Breuer, H.-P., Laine, E.-M., Vacchini, B.: Colloquium: Non-Markovian dynamics in open quantum systems. Rev. Mod. Phys. \textbf{88}, 021002 (2016).
\url{https://doi.org/10.1103/RevModPhys.88.021002}

\bibitem{bellomo2007non} 
Bellomo, B., Lo Franco, R., Compagno, G.: Non-Markovian effects on the dynamics of entanglement. Phys. Rev. Lett. \textbf{99}, 160502 (2007).
\url{https://doi.org/10.1103/PhysRevLett.99.160502}

\bibitem{utagi2020temporal} 
Utagi, S., Srikanth, R., Banerjee, S.:
Temporal self-similarity of quantum dynamical maps as a concept of memorylessness. Scientific Reports \textbf{10}, 20767 (2020). \url{https://doi.org/10.1038/s41598-020-72211-3}

 \bibitem{scaraniV} 
Scarani, V., Bechmann-Pasquinucci, H., Cerf, N. J., Dusek, M., Lutkenhaus, N., Peev, M.:
The security of practical quantum key distribution.
Rev. Mod. Phys. \textbf{81}, 1301 (2009).
\url{https://doi.org/10.1103/RevModPhys.81.1301}

 \bibitem{DW05} 
Devetak, I., Winter, A.: Distillation of secret key and entanglement from quantum states. Royal Society \textbf{461}, 2005.
\url{https://doi.org/10.1098/rspa.2004.1372}


\bibitem{sp09} 
Pironio, S., Acín, A., Brunner, N., Gisin, N., Massar, S., Scarani, V.:
Device-independent quantum key distribution secure against collective attacks. New Journal of Physics \textbf{11}, 045021 (2009).
\url{https://doi.org/10.1088/1367-2630/11/4/045021}

\bibitem{Hol73}Holevo, A. S.: Bounds for the Quantity of Information Transmitted by a Quantum Communication Channel. Problemy Peredachi Informatsii
(1973).\url{https://api.semanticscholar.org/CorpusID:118312737}

\bibitem{Hu2016} Hu, J. Y., Yu, B., Jing, M. Y., et al.: Experimental quantum secure direct communication with single photons. Light Science \& Applications \textbf{5}, e16144 (2016). \url{https://doi.org/10.1038/lsa.2016.144}

\bibitem{tang2012} 
Tang, J.-S., Li, C.-F., Li, Y.-L., Zou, X.-B., Guo, G.-C., Breuer, H.-P., Laine, E.-M., Piilo, J.:
Measuring non-Markovianity of processes with controllable system-environment interaction.
EPL \textbf{97}, 10002 (2012).
\url{https://doi.org/10.1209/0295-5075/97/10002}

\bibitem{passos2019} 
Passos, M., Obando, P., Balthazar, W., Paula, F., Huguenin, J., Sarandy, M.:
Non-Markovianity through quantum coherence in an all-optical setup. Optics Letters \textbf{44}, 2478-2481 (2019). \url{https://doi.org/10.1364/OL.44.002478}

\bibitem{kuhr2007ultrahigh} 
Kuhr, S., Gleyzes, S., Guerlin, C., Bernu, J., Hoff, U. B., Deléglise, S., Osnaghi, S., Brune, M., Raimond, J.-M., Haroche, S., et al.:
Ultrahigh finesse Fabry-Pérot superconducting resonator. Applied Physics Letters \textbf{90}, 16 (2007). \url{https://doi.org/10.1063/1.2724816}

\bibitem{milul2023superconducting} 
Milul, O., Guttel, B., Goldblatt, U., Hazanov, S., Joshi, L. M., Chausovsky, D., Kahn, N., Çiftyürek, E., Lafont, F., Rosenblum, S.:
Superconducting Cavity Qubit with Tens of Milliseconds Single-Photon Coherence Time.
PRX Quantum \textbf{4}, 030336 (2023). \url{http://doi.org/10.1103/PRXQuantum.4.030336}

\bibitem{zapatero2019} 
Zapatero, V., Curty, M.:
Long-distance device-independent quantum key distribution. Scientific Reports \textbf{9}, 17749 (2019). \url{https://doi.org/10.1038/s41598-019-53803-0}


\bibitem{wang2018} 
Wang, K.-H., Chen, S.-H., Lin, Y.-C., Li, C.-M.:
Non-Markovianity of photon dynamics in a birefringent crystal. Phys. Rev. A \textbf{98}, 043850 (2018). \url{https://doi.org/10.1103/PhysRevA.98.043850}


\end{thebibliography}
\end{document}